\documentclass{TTP_DSL2006}
\usepackage[pdftex]{graphicx}
\usepackage[intlimits]{amsmath}
\usepackage{amssymb}
\usepackage{exscale}
\usepackage{ifpdf}
\usepackage{epstopdf}

\begin{document}

\def\taub{\mbox{\boldmath $\tau$}}

\title{Recent progress in the thermodynamics of ferrotoroidic materials}

\author{Antoni Planes\inst{1}, Teresa Cast\'an\inst{1}, Avadh Saxena\inst{2} }

\institute{Departament d'Estructura i Constituents de la
  Mat\`eria, Facultat de F\'{\i}sica, Universitat de
  Barcelona, Diagonal 647, 08028 Barcelona, Catalonia
\and Theoretical Division, Los Alamos National Laboratory, Los Alamos, New Mexico 87545, USA}

\maketitle

\vspace{-3mm}
\sffamily
\begin{center}
\end{center}


\vspace{3mm} \hspace{-7.7mm} 
\rmfamily \normalsize \textbf{Abstract.} Recent theoretical and experimental progress on the study of ferrotoroidic materials is reviewed.
The basic field equations are first described and then the expressions for magnetic toroidal moment and toroidization are derived. Relevant materials and experimental observation of magnetic toroidal moment and toroidal domains are summarized next. The thermodynamics of such magnetic materials is discussed in detail with examples of ferrorotoidic phase transition studied using Landau modelling. Specifically, an example of application of Landau modelling to the study of toroidocaloric effect is also provided. Recent results of polar nanostructures with electrical toroidal moment are finally reviewed.


%
\section{1. Introduction}
According to the common point of view, ferroelastic, ferroelectric and ferromagnetic materials constitute the family of ferroic materials \cite{Wadhawan2000}. More recently ferrotoroidic materials have also been included in this family \cite{Schmid2001,Spaldin2008}. Ferrotoroidics describe materials where toroidal moments show cooperative long range order. Ferrotoroidic materials intrinsically belong to the class of multiferroic materials \cite{Khomskii2009}. Ferrotoroidicity spontaneously emerges at a phase transition from a paratoroidic to a ferrotoroidic phase in which both time and spatial inversion symmetries are simultaneously broken \cite{Schmid2001,Spaldin2008,Khomskii2009,Saxena2011}. The order parameter for this transition is toroidization.  Note that in ferroelectrics only the spatial inversion symmetry is broken whereas in ferromagnets only the time reversal symmetry is broken.  In ferroelastics neither symmetry is broken; only the rotational symmetry is broken \cite{Saxena2011}.

Ferrotoroidal order can be understood in terms of an ordering of magnetic-vortex like structures characterized by a toroidal (dipolar) moment.  This order is also related to (asymmetric) magnetoelectricity, i.e. $\alpha_{ij}\ne\alpha_{ji}$ \cite{Spaldin2008}.   In the present article we are mainly concerned with magnetic toroidal moments \cite{Schmid2001, Dubovik1986, Dubovik1990, Dubovik2000, Kopaev2009} as observed for instance, in LiCo(PO$_4$)$_3$ \cite{VanAken2007}.  Electrical toroidal moments can also exist in nanostructures such as polar dots \cite{Naumov2004,Prosan2008, Prosandeev2013} but not as a long range ordered state in bulk materials as no symmetry is broken.  
In Section 7 we will summarize recent results on nanoscale ferroelectric materials which exhibit electric toroidization as a consequence of dipolar vortex formation.

\begin{figure}[ht]
\centering
\includegraphics[scale=0.61]{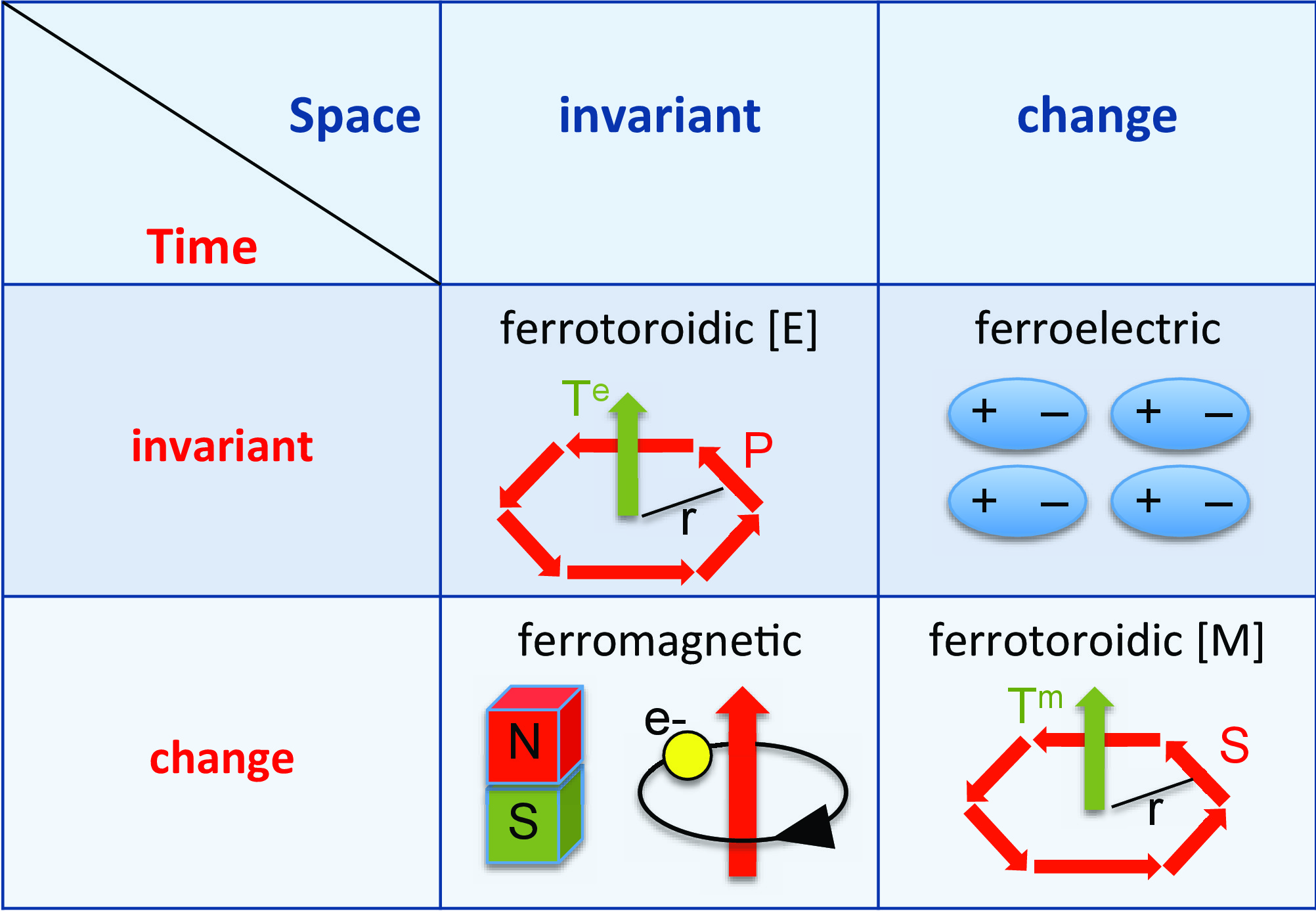}
\caption{Symmetry properties of the four vectorial ferroic orders.}
\label{Fig0}
\end{figure}

Figure \ref{Fig0} shows the symmetry properties of the four ferroic vector order parameters.    Polarization is a polar vector, magnetization is an axial vector (it contains a sense of time), and toroidization is an axio-polar vector.  Note that strain is a (second rank polar) tensor order parameter.  In addition, there are physical properties described by a second rank axial tensor such as magnetogyration \cite{Lines1981,Zheludev1985} likely present in the spin-half antiferromagnet potassium hyperoxide, KO$_2$, as well as in CdS, (Ga$_x$In$_{1-x}$)$_2$Se$_3$, Pb$_5$Ge$_3$O$_{11}$ and Bi$_{12}$GeO$_{20}$.

Therefore, the table can be conceivably generalized to include {\it tensor ferroics}.  The  two entries in the left column would be strain (second rank polar tensor) and magnetogyration (second rank axial tensor) but at present the tensor analogs of polarization and toroidization that would complete the table have not been properly identified. Note that in principle this idea could be generalized further to third (and higher) rank tensor ferroic properties.  We intend to present these results elsewhere in the near future.

Any ferroic order is usually accompanied by domain walls \cite{Wadhawan2000}.  Indeed, ferrotoroidic domain walls have been observed in LiCo(PO$_4$)$_3$ using nonlinear optics, i.e. second harmonic generation.  Litvin has provided a symmetry based classification of such domains \cite{Litvin08} as well as ferrotoroidal crystals \cite{LitvinActa, Litvin14}.  Another material,  BCG (Ba$_2$CoGe$_2$O$_7$), also exhibits spontaneous toroidal moments \cite{Toledano2011}.  Similarly MnTiO$_3$ thin films show ferrotoroidic ordering \cite{Toyosaki2008}.  Multiple ferrotoroidic phase transitions have been studied in Ni-Br and Ni-I boracites \cite{Kopaev2009, Sannikov1998}.  Interestingly, some quasi-one dimensional materials such as pyroxenes \cite{Pyroxene} also exhibit ferrotoroidic behaviour.


\section{2. Basic field equations\label{basic}}

We will introduce here the toroidic moment and toroidization following ideas published by Dubovik et al. in Ref. \cite{Dubovik2000}. Assume distributions of charges $\rho ({\bf r})$ and currents ${\bf j ({\bf r})}$ localized in a given region of space. These distributions create electric and magnetic fields that satify Maxwell equations. In the presence of matter, within  the continuum dipolar approximation, these equations are usually expressed as \cite{Jackson1974}:
\begin{eqnarray}
\nabla \times {\bf H} - \frac{\partial {\bf D}}{\partial t} & = & {\bf j},  \label{Max1}\\
\nabla \cdot {\bf D} & = & \rho, \label{Max2}\\
\nabla \times {\bf E} + \frac{\partial {\bf B}}{\partial t} & = & 0, \label{Max3}\\
\nabla \cdot {\bf B} & = & 0. \label{Max4}
\end{eqnarray}
The electric displacement, ${\bf D}$, and the magnetic field, ${\bf H}$, are defined as:
\begin{eqnarray}
{\bf D} & = & \varepsilon_0 {\bf E} + {\bf P}, \\
{\bf H} & = & \frac{1}{\mu_0} {\bf B} - {\bf M},
\end{eqnarray}
where ${\bf P}$ and ${\bf M}$ are the electric and magnetic polarizations of the medium, and $\varepsilon_0$ and $\mu_0$ the electric permittivity and the magnetic  permeability of free space. Electric polarization (or simply polarization) and magnetic polarization (or magnetization) are introduced as volume densities of the electric and magnetic moments which are defined from a multipole expansion far from charge and current distribution of the electric scalar potential, $\varphi$, and the magnetic vector potential, ${\bf A}$, at dipolar order respectively \cite{Raab2004}. These potentials are defined from eqs. (\ref{Max1}) and (\ref{Max4}) as:
\begin{eqnarray}
{\bf E} & = & - \nabla \varphi - \dot{\bf A}, \\
{\bf B} & = & \nabla \times {\bf A}.
\end{eqnarray}
We will see that the dipolar approximation is not sufficient for some peculiar (electric and magnetic) moment configurations. In this case, higher order terms in the expansion must be taken into account.

\begin{figure}[ht]
\centering
\includegraphics[scale=0.5]{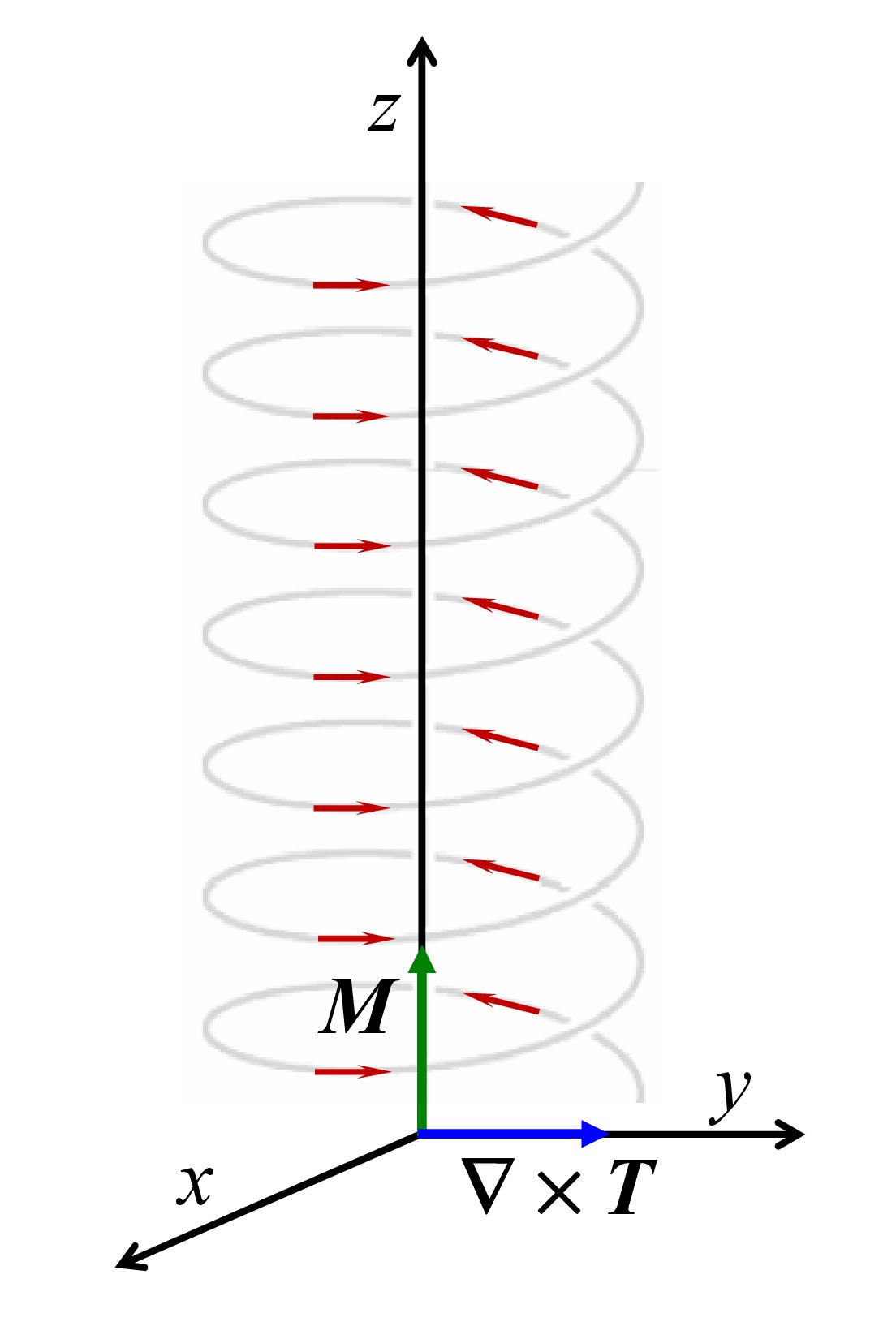}
\caption{Spiral-like configuration of moments along the surface of a cylinder.}
\label{Fig2}
\end{figure}

Now let us assume a situation where the configuration of magnetic or electric moments is spiral-like. This is illustrated in Fig. \ref{Fig2}. For this configuration, the magnetization (or polarization) is along the $z$-axis, while the projection on the $xy$-plane, which has a circle-like configuration, is zero. This circle-like configuration can be understood as being originated by a toroidal configuration of loop-currents or electric dipoles in the magnetic and electric cases respectively. For this kind of configurations ${\bf M}$ or ${\bf P}$ alone do not provide enough information about the ordering of magnetic/electric moments. Irene A. Beardsley \cite{Beardsley1989} noticed that in this case an arbitrary amount of a divergence-free magnetization/polarization distribution can be added to ${\bf M}$ or ${\bf P}$ without affecting the external field created by the distribution of magnetic/electric moments. This divergence-free term can be written as the curl of some vector that characterizes the circle-like configuration of magnetic/electric moments in the $xy$-plane. In the magnetic case, this vector is the magnetic toroidization, $\nabla \times {\bf T}^M$, while in the electric case it is the electric toroidization, $\nabla \times {\bf T}^E$. Note that the existence of magnetic toroidization implies that both, time-reversal and spatial-inversion symmetries are broken. Therefore, magnetic toroidization is represented by an axiopolar (or time-odd polar) vector. In contrast, no broken symmetry is associated with electric toroidization. As we will discuss later in Section 7, these toroidizations are related to the moment of the distributions of magnetic and electric moments respectively. In practice, this means that the magnetization ${\bf M}$ must be replaced by ${\bf M} + \nabla \times {\bf T}^M$, while in the electric case, the electric polarization ${\bf P}$ must be replaced by ${\bf P} +  \nabla \times {\bf T}^E$.

When electric and magnetic toroidizations are taken into account, the new macroscopic Maxwell equations take the same formal expressions as the standard ones after the following redefinition of the fields.
\begin{eqnarray}
{\bf D} & \rightarrow & {\bf {\widehat D}} = \varepsilon_0 {\bf E} + {\bf P} + \nabla \times {\bf T}^E = \varepsilon_0 {\bf E} + {\bf \widehat{P}}, \\
{\bf H} & \rightarrow & {\bf {\widehat H}} = \frac{1}{\mu_0} {\bf B} - {\bf M} - \nabla \times {\bf T}^M = \frac{1}{\mu_0}{\bf B} - {\bf \widehat{M}}.
\end{eqnarray}
Therefore, the new macroscopic Maxwell equations read:
\begin{eqnarray}
\nabla \times {\bf \widehat{H}} - \frac{\partial {\bf \widehat{D}}}{\partial t} & = & {\bf j},  \label{MaxT1}\\
\nabla \cdot {\bf \widehat{D}} & = & \rho, \label{MaxT2}\\
\nabla \times {\bf E} + \frac{\partial {\bf B}}{\partial t} & = & 0, \label{MaxT3}\\
\nabla \cdot {\bf B} & = & 0. \label{MaxT4}
\end{eqnarray}
%
%
%

The energy density accounting for the interaction of the generalized polarization and magnetization with external fields are given by ${\bf E} \cdot {\bf \widehat{P}}$ and ${\bf B} \cdot {\bf \widehat{M}}$ respectively. Hence, the coupling energies of the field with electric and magnetic toroidizations are  $\int{\bf E} \cdot [\nabla \times {\bf T}^E]  d^3r$ and $\int{\bf B} \cdot [\nabla \times {\bf T}^M] \;\; d^3r$ respectively. These terms can be written in the form:
\begin{eqnarray}
\int{\bf E} \cdot [\nabla \times {\bf T}^E]  d^3r & = & \int \nabla \cdot ({\bf T}^E \times {\bf E}) d^3r + \int (\nabla \times {\bf E}) \cdot {\bf T}^E d^3r, \label{for1}\\
\int{\bf B} \cdot [\nabla \times {\bf T}^M]  d^3r & = & \int \nabla \cdot ({\bf T}^M \times {\bf B}) d^3r + \int (\nabla \times {\bf B}) \cdot {\bf T}^M d^3r, \label{for2}
\end{eqnarray}
where the first terms on the right-hand  sides of both eq. (\ref{for1}) and eq. (\ref{for2}) are zero taking into account the Gauss theorem. In the electric case, $\nabla \times E = 0$, and thus this energy vanishes. In the magnetic case, it is given by $\int (\nabla \times {\bf B}) \cdot {\bf T}^M d^3r$. Therefore, this shows that the conjugate field of the magnetic toroidization is $\nabla \times {\bf B}$.

\begin{figure}[ht]
\centering
\includegraphics[scale=0.55]{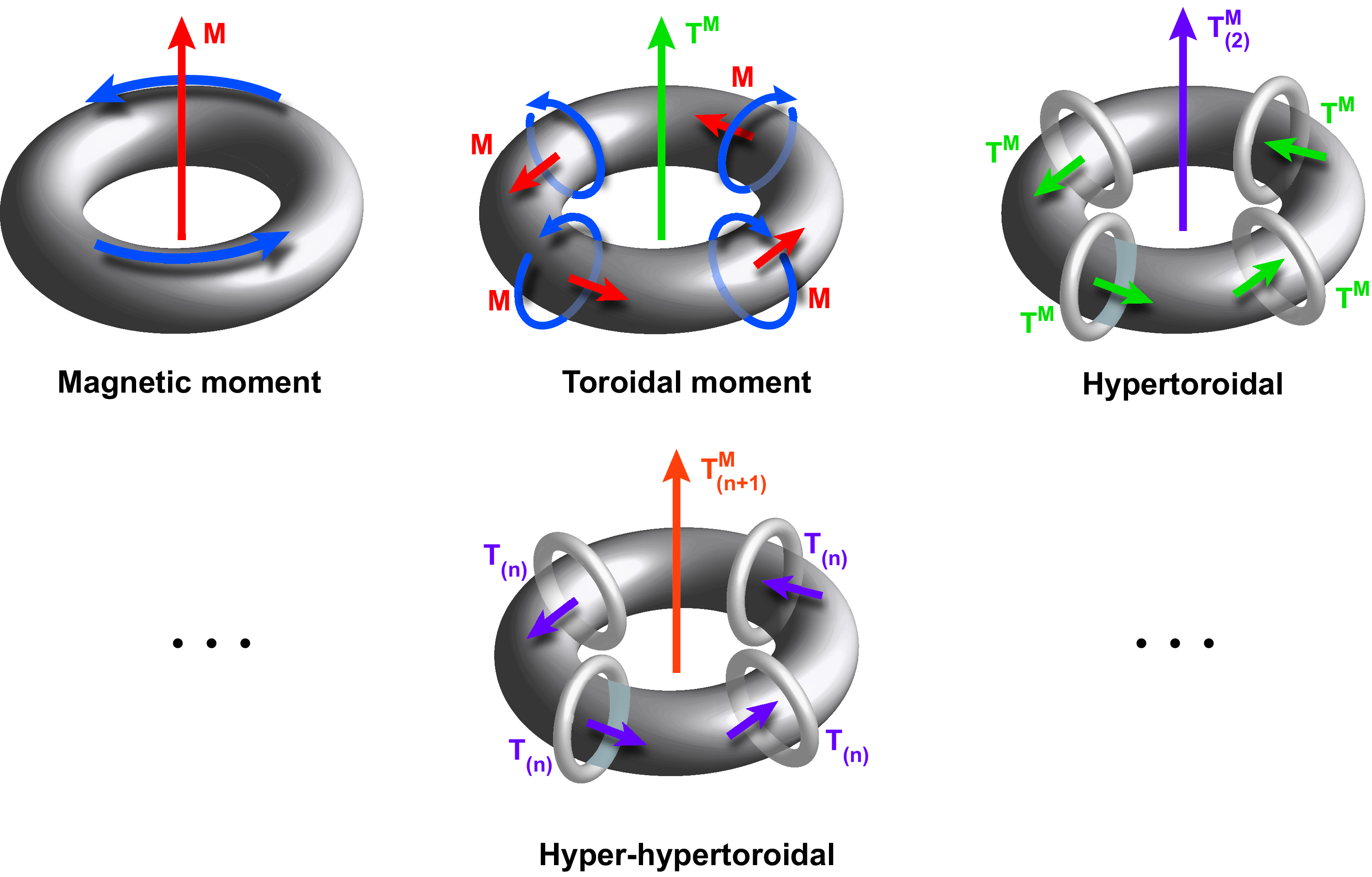}
\caption{Magnetic moment, toroidal moment and the generation of successive hypertoroidal moments.}
\label{Fig3}
\end{figure}

We can now generalize the ideas discussed above and assume systems with spiral-like configurations of toroidal moments (either electric or magnetic) arising from toroidal configurations of electric or magnetic moments. This kind of double-vortex configuration should be characterized by divergence-free vectors expressed as the curl of higher order toroidal moments.  These moments  are denoted as hypertoroidal moments \cite{Prosandeev2009}.  In the case of magnetism, for instance, this means that in the macroscopic Maxwell equations the toroidization  ${\bf T}^M$, should be replaced by:
\begin{eqnarray}
{\bf T}^M \rightarrow {\bf T}^M  + \nabla \times {\bf T}^M_{(2)},
\end{eqnarray}
where ${\bf T}^M = {\bf T}^M_{(1)}$ is the magnetic first-order toroidization and ${\bf T}^M_{(2)}$ is the magnetic second-order toroidization (or hypertoroidization). Hence, magnetization should be replaced by:
\begin{eqnarray}
{\bf M} \rightarrow {\bf \widehat{M}}_{(2)} = {\bf M} + \nabla \times {\bf T}_{(1)}^M  + \nabla \times (\nabla \times {\bf T}^M_{(2)}).
\end{eqnarray}
Of course this idea can be formally generalized to any order by defining higher order hypertoroidal moments (see Fig. \ref{Fig3}). At order $n$ we will have:
\begin{eqnarray}
{\bf \widehat{M}}_{(n)} = {\bf M} + \nabla \times {\bf T}_{(1)}^M  + \nabla \times \nabla \times {\bf T}^M_{(2)} + ...+ \nabla \times ....\times \nabla \times {\bf T}^M_{(n)}.
\end{eqnarray}
Indeed, one can proceed similarly in the electric case. Therefore, Maxwell equations at the $n$th order, similar to equations (\ref{MaxT1}-\ref{MaxT4}),  can be established by defining the following fields:
\begin{eqnarray}
{\bf \widehat{D}}_{(n)} & = & \varepsilon_0 {\bf E} + {\bf \widehat{P}}_{(n)}, \\
{\bf \widehat{H}}_{(n)} & = & \frac{1}{\mu_0} {\bf B} - {\bf \widehat{M}}_{(n)}.
\end{eqnarray}

With similar arguments as those given above, it is easy to show that the field conjugated to the $n$th order magnetic hypertoroidization is, $\nabla \times ....\times \nabla \times {\bf B}$.

\section{3. The magnetic toroidal moment and toroidization \label{toroidal-moment}}

Similar to polarization and magnetization, electric and magnetic toroidizations are defined in the continuum approximation as volume densities of electric and magnetic toroidal moments respectively. Since the symmetries associated with electric toroidization are trivial (no change of sign is expected either under spatial inversion or under time reversal) no phase transition to an electric toroidal phase should be envisaged. Actually, this is consistent with the fact that toroidal moment associated with electric moment vorticity is not expected to occur in the thermodynamic limit \cite{comment1}. From the discussion in the preceding section, it seems intuitively reasonable to foresee that the toroidal moment should be related to the moment of the distribution of magnetic moments. This is what we will discuss in the rest of this section where we will introduce the toroidal moment based on the multipolar expansion beyond the dipolar approximation. We will also introduce a second definition based on symmetry considerations which is of interest from a more macroscopic thermodynamic point of view.


In the former case we consider a finite distribution of steady currents ${\bf J}({\bf r})$. The vector potential of this distribution is given by
\begin{eqnarray}
{\bf A}({\bf R}) = \frac{\mu_0}{4 \pi}\int_V \frac{{\bf J}({\bf r})}{|{\bf R} - {\bf r}|} dv,
\end{eqnarray}
where ${\bf R}$ is the vector position of a point $P$, ${\bf r}$ the vector position of the volume element $dv$ and $V$ the volume of the distribution, and the Coulomb gauge ($\nabla \cdot {\bf A} = 0$) has been assumed. The multipole expansion of ${\bf A}({\bf R})$ (about ${\bf r} = 0$) takes the form \cite{Raab2004},
%
%
\begin{eqnarray}
{\bf A}({\bf R}) = \frac{\mu_0}{4 \pi} \sum_{n = 0}^\infty \frac{(-1)^nç}{n!} \int_V {\bf J}({\bf r})[{\bf r} \cdot \nabla]^n \left(\frac{1}{R}\right) dv.
\end{eqnarray}

It is easy to see that the zeroth-order term vanishes for a steady current distribution for which the continuity equation yields $\nabla \cdot {\bf J} = 0$.  The first order term in the expansion is the dipolar term that can be expressed as,
\begin{eqnarray}
{\bf A}^{(1)} = - {\bf m} \times \nabla \frac{1}{R} = \frac{{\bf m} \times {\bf R}}{R^3} ,
\end{eqnarray}
where ${\bf m}$ is the magnetic dipolar moment defined as,
\begin{eqnarray}
{\bf m} = \frac{1}{2} \int_V ({\bf r} \times {\bf J}) dv.
\end{eqnarray}
The next term can be expressed as the sum of magnetic quadrupolar and toroidal contributions. The quadrupolar part is given by,
\begin{eqnarray}
(A_{quad})_i^{(2)} = -\varepsilon_{ijk} q_{kl} \nabla_i \nabla_l \frac{1}{R}
\end{eqnarray}
%
%
where $\varepsilon_{ijk}$ is the Levi-Civita symbol and $q_{kl}$ is the magnetic quadrupolar moment given by,
\begin{eqnarray}
q_{ij} = \frac{2}{3} \int_V ({\bf r} \times {\bf J})_i r_j dv,
\end{eqnarray}
which is a traceless symmetric tensor. The toroidal term is given by,
\begin{eqnarray}
{\bf A}_{tor}^{(2)} = \nabla ({\bf t} \cdot \nabla) \frac{1}{R} + {\bf t} \delta ({\bf R}),
\end{eqnarray}
where ${\bf t}$ is the toroidal moment that can be expressed as
\begin{eqnarray}
{\bf t} = \frac{1}{4} \int_v {\bf r} \times [{\bf r} \times {\bf J}({\bf r})] dv .
\label{t-moment}
\end{eqnarray}
This pseudovector represents the dual antisymmetric part of the complete tensor which appears in the second order term of the multipole expansion. Defining ${\bf m}({\bf r}) = \frac{1}{2}[{\bf r} \times {\bf J}({\bf r})]$ as the distribution of magnetic moments, the toroidal moment can be written as,
\begin{eqnarray}
{\bf t} = \frac{1}{2} \int_v [{\bf r} \times {\bf m}({\bf r})] dv,
\end{eqnarray}
which indicates that the toroidal moment can be understood as the moment of the distribution of magnetic moments (see Fig. 3b).

For a discrete distribution of $N$ point charges $q_{\alpha}$ of mass $m_{\alpha}$ localized at positions ${\bf r}_{\alpha}$ with velocities ${\bf u}_{\alpha}$, the current density can be written as
\begin{eqnarray}
{\bf J ({\bf r})} =  \sum_{\alpha=1}^N q_{\alpha} {\bf u}_{\alpha} \delta({\bf r} - {\bf r}_{\alpha}).
\end{eqnarray}
The magnetic moment then takes the form:
\begin{eqnarray}
{\bf m} = \frac{1}{2} \sum_{\alpha =1}^N q_{\alpha} {\bf r}_{\alpha} \times {\bf u}_{\alpha} ,
=  \sum_{\alpha =1}^N {\bf m}_{\alpha}
\end{eqnarray}
where ${\bf m}_{\alpha}$ is the magnetic moment of charge $\alpha$.
Similarly, the toroidal moment of interest here can be written as,
\begin{eqnarray}
{\bf t} = \frac{1}{4} \sum_{\alpha = 1}^N q_{\alpha} ({\bf r}_{\alpha} \times [{\bf r}_{\alpha} \times {\bf u}_{\alpha}]) = \frac{1}{2} \sum_{\alpha = 1}^N [{\bf r}_{\alpha} \times {\bf m}_{\alpha}].
\end{eqnarray}


For a system consisting of a distribution of $N$ spins ${\bf s}_{\alpha}$ localized at positions ${\bf r}_{\alpha}$, ${\bf m}_{\alpha} = g \mu_0 {\bf s}_{\alpha}$,
%
%
%
%
where $g$ is the gyromagnetic ratio and $\mu_B$ the Bohr magneton.
%
%
Therefore, the corresponding toroidal moment is given by
\begin{eqnarray}
{\bf t}_{\alpha} = \frac{1}{2}g \mu_0 \sum_{\alpha =1}^N ({\bf r}_{\alpha} \times {\bf s}_{\alpha}).
\end{eqnarray}
%
%
%

It is worth noting that treatment similar to the one discussed above can be developed from the multipolar expansion of the electric scalar potential $\phi$. In particular, from the first order term the electric dipolar moment can be defined as
\begin{eqnarray}
{\bf p} = \int_V {\bf r} \rho({\bf r})dv.
\end{eqnarray}

Once the toroidal moment is introduced, magnetic toroidization, hereafter simply denoted as toroidization, ${\bf T}$,  is defined as the volume density of toroidal moments. That is $ {\bf T} = d{\bf t} / dv$. We have already seen that its corresponding conjugated field is $\nabla \times {\bf B}$ and hence, the energy density of a distribution of toroidal moments characterized by a toroidization ${\bf T}$ is given by $E = - {\bf T} \cdot (\nabla \times {\bf B})$. Therefore, a net toroidization might be induced by means of a current density ${\bf J} = \nabla \times {\bf B}$. Schmid \cite{Schmid2001} noticed that reversing toroidal dipoles by means of such a field in order to modify toroidization appears unfeasible since this would require the action of coherent circular currents of very small size (comparable to the unit cell of the crystal).

The observation of toroidization in the absence of applied fields indicates the existence of long range order associated with toroidal moments. This long range order is usually denoted as ferrotoroidic order and should be related to some kind of coupling between toroidal moments. Taking into account the basic symmetries of the toroidal moment, the occurrence of ferrotoroidic order supposes the simultaneous breaking of spatial inversion and time reversal symmetries.

Materials with toroidal moments are expected to intrinsically display magnetoelectric coupling. In these systems, an applied magnetic field breaks inversion symmetry and thus induces polarization. On its turn, an applied electric field breaks time reversal symmetry and induces magnetization. Therefore, we expect that these materials respond to applied electric and magnetic fields according to the following equations,
\begin{eqnarray}
{\bf P} & = & {\overline{\overline{\chi}}}_{\;e} \;{\bf E} + {\overline{\overline{\alpha}}}^T \; {\bf B},  \\
{\bf M} & = & {\overline{\overline{\alpha}}}^T \;{\bf E} + {\overline{\overline{\chi}}}_{\;m} \; {\bf B} ,
\end{eqnarray}
where ${\overline{\overline{\chi}}}_{\;e}$ and ${\overline{\overline{\chi}}}_{\;m}$ are respectively, electric and magnetic susceptibility tensors, and ${\overline{\overline{\alpha}}}^T$ is the magnetoelectric tensor (all are rank-2 tensors). From a thermodynamic point of view, if the free energy of the system is ${\cal F}$, these polarizations and magnetizations should be expressed as $- \partial {\cal F} / \partial {\bf E}$ and $- \partial {\cal F} / \partial {\bf B}$, respectively. Therefore, we expect that the free energy of a magnetoelectric term is of the type ${\cal F}_{m-e} = - {\bf E} \;  {\overline{\overline{\alpha}}}^T \; {\bf B}$ (or $ - \alpha_{ij}^T E_i B_j$, in coordinate notation). The decomposition of this magnetoelectric term into pseudoscalar, vector, and symmetric traceless terms enables one to express ${\cal F}_{m-e}$ in the form (see Ref. \cite{Spaldin2008}),
\begin{eqnarray}
{\cal F}_{m-e} \sim -  {\bf E} \cdot {\bf B} - {\bf T}' \cdot [{\bf E} \times {\bf B}] - Q_{ij} [E_i B_j + E_j B_i] ,
\end{eqnarray}
where ${\bf T}'$ is a vector with the same symmetry properties as toroidal moment and toroidization. Identifying this vector with toroidization ${\bf T}$ supposes that its conjugated field is ${\bf G} = {\bf E} \times {\bf B}$. This assumption is in agreement with recent experiments that showed that toroiodal moments can be controlled by this field \cite{Baum2013}. Therefore, polarization, ${\bf P}_t$, and magnetization, ${\bf M}_t$,  intrinsically associated with the energy term $- {\bf G} \cdot {\bf T}$, induced, respectively, under application of electric and magnetic fields, can be expressed as
\begin{eqnarray}
{\bf P}_t & = & - \frac{\partial {\bf T} \cdot [{\bf E} \times {\bf B}]}{\partial {\bf E}} = {\bf B} \times {\bf T}, \label{Pt}\\
{\bf M}_t & = & - \frac{\partial {\bf T} \cdot [{\bf E} \times {\bf B}]}{\partial {\bf B}} = {\bf T} \times {\bf E}, \label{Mt}
\end{eqnarray}
where we have taken into account that ${\bf T} \cdot [{\bf E} \times {\bf B}] = {\bf B} \cdot [{\bf T} \times {\bf E}] = {\bf E} \cdot [{\bf B} \times {\bf T}]$.

In the far-field approximation, from multipolar expansions of electric (scalar) and vector potentials, the toroidal field ${\bf G}$ can be expressed as,
\begin{eqnarray}
{\bf G} = {\bf E} \times {\bf B} =  A ({\bf p} \times {\bf m}) + B ({\bf p} \times {\bf r}) + C ({\bf r} \times {\bf m}), \label{Toroidalfield}
\end{eqnarray}
where, ${\bf p}$ and ${\bf m}$ are electric and magnetic dipole moments, and the coefficients $A$, $B$, and $C$ are coefficients that decay with distance $r$ as $r^{-6}$, $r^{-7}$ and $r^{-7}$ respectively.
%
%
Taking into account eq. (\ref{Toroidalfield}), it is worth pointing out that if ${\bf p}$ and ${\bf m}$ are parallel, then ${\bf G} = 0$ as expected. On the other hand, ${\bf G}$ has maximum strength when ${\bf p}$ and ${\bf m}$ are perpendicular.

Taking ${\bf G}$ as the field conjugated to toroidization, suggests the following alternative definition of the toroidal moment:
\begin{eqnarray}
{\bf t} = \frac{\mu_0}{4 \pi} ({\bf p} \times {\bf m}).
\end{eqnarray}
This definition neglects residual terms in eq. (\ref{Toroidalfield}) associated with the magnetic and electric moments. The choice is supported by the fact that these terms decay (with distance) faster than the magneto-toroidal one.

It is worth noticing that this definition of the toroidal moment is not strictly equivalent to the definition in eq. (\ref{t-moment}) resulting from the second order term in the multipole expansion of the vector potential. Nevertheless, it is, in fact, expected to provide a good measure of the toroidal moment in systems which are simultaneously ferroelectric and ferromagnetic \cite{Sawada2005}.
In these systems the coupling of ${\bf t}$ (or the toroidization obtained as the volume density of this toroidal moment) to ${\bf G}$ leads to a magnetoelectric response similar to that of magnetic ferrotoroidics \cite{Spaldin2008}.
In general, however, a non-zero toroidal moment should be possible even in antiferroelectric and
antiferromagnetic systems. Indeed, resonant x-ray diffraction observations of orbital currents in CuO provide direct evidence of antiferrotoroidic ordering \cite{antitoroidic}. 
Actually, these situations can only be considered when the standard definition (arising from the multipole expansion) of the toroidal moment is taken into
account.   Note that a multipole expansion including the toroidal moment has been considered in \cite{Gongora2006}.

%
%

%
%

%
%

\section{4. Materials and relevant experimental results}

At present direct measurements of toroidization or toroidal moment seem very difficult. Present experimental techniques can only detect magnetization and magnetic moment, for instance from polarized neutron scattering or Lorentz microscopy. In principle, these techniques should be able to detect specific arrangements of magnetic moments characterized by toroidal moments that might order to yield net toroidization and thus, ferrotoroidal order. In practice, this appears to be unfeasible.

Indirectly ferrotoroidal order can be inferred from an asymmetric magnetoelectric response. Therefore, this needs measurement and analysis of the appropiate magnetoelectric tensor components. Sannikov \cite{Sannikov2007} has argued that observation of $\alpha_{ij} \neq \alpha_{ji}$ is an indication of possible ferrotoroidic order. It is however important to take into account that this condition is not sufficient to justify the occurrence of toroidization. Asymmetric behaviour of the magnetoelectric tensor has been reported for some boracites (G$_2$ phase of Co-I and Ni-Cl boracites). It has been reported also for some oxides such as Ga$_{2-x}$Fe$_x$O$_3$ and Cr$_2$O$_3$ \cite{Popov1998,Popov1999}.

Visualization of ferrotoroidal order requires an experimental technique that is sensitive to both space inversion and time reversal broken symmetries which is the inherent feature associated with ferrotoroidal order. As shown by Van Aken and co-authors \cite{VanAken2007} non-linear optics offers this possibility. These authors used optical second harmonic generation (SHG) to resolve ferrotoroidal domains in LiCoPO$_4$. Similar experiments were already carried out some years before \cite{Fiebig1994,Fiebig1995} but were much less conclusive in relation to the existence of ferrotoroidal domains. In this technique, electromagnetic light field ${\bf E}(\omega)$  of given frequency is incident on a crystal and induces a polarization at double the frequency which acts as a wave source. The symmetry affects the corresponding susceptibility. This means that the second harmonic generation light from domains with opposite order should have a phase shift of 180$^\circ$.

LiCoPO$_4$ crystallizes in the orthorhombic $Pnma$ olivine structure \cite{Molenda2006,Szewczyk2011}. It displays unique properties including large linear magnetoelectric effect and large Li-ionic conductivity. Co$^{2+}$ ions belonging to (100) Co-O layers carry the magnetic moments that are strongly coupled by superexchange Co-O-Co interactions. Layers, however, are only weakly coupled by higher order interactions. Thus, the system behaves as a magnetic 2-$d$ system to a very good approximation.

Antiferromagnetic order in the material occurs below a N\'eel temperature $T_N$ = 21.4 K. Due to large magnetocrystalline anisotropy the magnetic moments are confined to  directions lying within $b$-$c$ planes, approximately 4.6$^{\circ}$ away from the $b$ axis. Actually, the Co magnetic moments are not completely compensated, and the system shows a small net magnetic moment which, in fact, is not consistent with the orthorhombic symmetry and can only be understood assuming a small monoclinic distortion. Interestingly, the monoclinic symmetry allows for a non-zero dielectric polarization and a non-zero toroidal moment (along the pseudo-orthorhombic $a$-axis) to occur.

Van Aken et al. SHG experiments \cite{VanAken2007} detected four different domain states. It was shown later from symmetry considerations \cite{Schmid2008} that the four domains are  equivalent with different orientations of the net magnetic moment. Thus they carry toroidal moments with signs and directions mutually coupled. The fact that the SHG signal intensity is observed to disappear precisely at the N\'eel temperature, corroborates the coupling between magnetic and toroidal order parameters.

In addition to Ba$_2$CoGe$_2$O$_7$ \cite{Toledano2011} and MnTiO$_3$ thin films \cite{Toyosaki2008} toroidal moments have been considered in BiFeO$_3$ and related multiferroics \cite{Gorbat1994, Popov2001, Tokura2007}.  Another candidate material is the magnetoelectric MnPS$_3$ as indicated by neutron polarimetry \cite{Ressouche2010}.  Based on {\it an initio} calculations the olivine Li$_4$MnFeCoNiP$_4$O$_{16}$ is possibly a ferrotoroidic material \cite{Feng2009}. Note that apart from quasi-one dimensional materials called pyroxenes \cite{Pyroxene} the toroidal moment in the molecular context is also of interest, e.g. in dysprosium triangle based systems \cite{Lin2012}.  In a related context a physical realization of toroidal order is an interacting system of disks with a triangle of spins on each disk \cite{Harris2010}.

It is worth noting that the existence of a true long range ordered ferrotoroidic phase has only been established in a sufficiently reliable way in very few cases and, perhaps the clearer evidence has been provided by Van Aken et al. SHG results \cite{VanAken2007}. In other cases, only indirect results suggest the existence of such a phase. The fact that interaction between toroidal moments is very weak as indicated by the short range dipolar interaction (see eq. (\ref{Toroidalfield})), suggests that any small amount of disorder in the material is enough to yield a toroidal glassy state, which represents a frozen state with local order only \cite{Castan2014}. The existence of toroidal glass in Ni$_{0.4}$Mn$_{0.6}$TiO$_{3}$ has been foreseen from the behaviour of the magnetoelectric response which was observed to strongly depend on cooling history \cite{Yamaguchi2013}. This is indeed a very interesting result suggesting that materials which are candidates to display ferrotoroidal order should also be analysed within this point of view.  In fact, possible observation of toroidal glass completes the quartet of ferroic glasses, namely spin glass, relaxor ferroelectrics and strain glass
\cite{Ren2009}.

\section{5. Thermodynamics \label{Thermodynamics}}

We consider a macroscopic body where the vector ferroic properties, namely polarization, ${\bf P}$, magnetization, ${\bf M}$, and toroidization, ${\bf T}$, coexist.
Only part of the polarization and magnetization will be assumed to be intrinsic, thus originating from {\it preexisting} electric and magnetic moments. The remaining part will arise from the toroidization originating from magnetic toroidal momemts in the presence of external magnetic and electric fields respectively. That is,
\begin{eqnarray}
{\bf P} & = & {\bf P}_i + {\bf P}_t, \label{P=Pi+Pt}\\
{\bf M} & = & {\bf M}_i + {\bf M}_t. \label{M=Mi+Mt}
\end{eqnarray}
For this kind of closed systems, the fundamental thermodynamic equation reads
\begin{eqnarray}
dU = \tau dS + {\bf E} \cdot d {\bf P} + {\bf B} \cdot d {\bf M}, 
\end{eqnarray}
where $U$ is the internal energy density, $S$ the entropy density and $\tau$ the temperature.
%
%

Taking into account eq. (\ref{Pt}), the term ${\bf E} \cdot d{\bf P}$ can be expressed as,
\begin{eqnarray}
{\bf E} \cdot d{\bf P} = {\bf E} \cdot d{\bf P}_i + {\bf E} \cdot d{\bf P}_t = {\bf E} \cdot d{\bf P}_i + {\bf G} \cdot d{\bf T} + {\bf T} \cdot [{\bf E} \times d{\bf B}].
\end{eqnarray}
Similarly, using eq. (\ref{Mt}), the term ${\bf B} \cdot {\bf dM}$ can be expressed as,
\begin{eqnarray}
{\bf B} \cdot d{\bf M} = {\bf B} \cdot d{\bf M}_i + {\bf B} \cdot d{\bf M}_t = {\bf B} \cdot d {\bf M}_i + {\bf G} \cdot d {\bf T} + {\bf T} \cdot [d{\bf E} \times {\bf B}].
\end{eqnarray}
Therefore,
\begin{eqnarray}
dU = \tau dS + {\bf E} \cdot d {\bf P}_i + {\bf B} \cdot d {\bf M}_i + {\bf G} \cdot d {\bf T} + d({\bf G} \cdot {\bf T}).
\end{eqnarray}

%
%

Helmholtz, ${\cal F}$, and Gibbs, ${\cal G}$, free energies are defined as follows, 
\begin{eqnarray}
{\cal F} & = & U - \tau S \\
{\cal G} & = & {\cal F} - {\bf E} \cdot {\bf P} - {\bf B} \cdot {\bf M} = {\cal F} - {\bf E} \cdot {\bf P}_i - {\bf B} \cdot {\bf M}_i - 2 {\bf G} \cdot {\bf T}.
\end{eqnarray}
Their differential expressions are,
\begin{eqnarray}
d{\cal F} & = & - S d\tau + {\bf E} \cdot d {\bf P} + {\bf B} \cdot d {\bf M},\\
d{\cal G} & = & - S d\tau - {\bf P} \cdot d{\bf E} - {\bf M} \cdot d{\bf B},
\end{eqnarray}
which can be alternatively expressed as,
\begin{eqnarray}
d{\cal F} & = & - S d\tau + {\bf E} \cdot d {\bf P}_i + {\bf B} \cdot d {\bf M}_i + {\bf G} \cdot d {\bf T} + d({\bf G} \cdot {\bf T}),\\
d{\cal G} & = & - S d\tau - {\bf P}_i \cdot d{\bf E} - {\bf M}_i \cdot d{\bf B} - {\bf T} \cdot d{\bf G}.
\end{eqnarray}
Note that this expression suggests that we can assume that the three ferroic properties, polarization, magnetization, and toroidization can be assumed as independent (vector) quantities thermodynamically conjugated to the electric, magnetic and toroidal fields, respectively.

The response of the system to applied electric and magnetic fields is given by the generalized susceptibility,
\begin{eqnarray}
\xi =
\left(
  \begin{array}{cc}
    \frac{\partial^2 {\cal G}}{\partial {\bf E}^2} & \frac{\partial^2 {\cal G}}{\partial {\bf B}\partial{\bf E}} \\
    \frac{\partial^2 {\cal G}}{\partial {\bf E} \partial {\bf B}} & \frac{\partial^2 {\cal G}}{\partial {\bf B}^2} \\
  \end{array}
\right) =
  - \left(
    \begin{array}{cc}
      \frac{\partial {\bf P}}{\partial {\bf E}}  & \frac{\partial {\bf P}}{\partial {\bf B}} \\
      \frac{\partial {\bf M}}{\partial {\bf E}} & \frac{\partial {\bf M}}{\partial {\bf B}} \\
    \end{array}\right) =
    - \left(
  \begin{array}{cc}
    {\overline{\overline \chi}}_e & {\overline{\overline \alpha}} \\
    {\overline{\overline \alpha}}^T & {\overline{\overline \chi}}_m \\
  \end{array}
  \right).
\end{eqnarray}
Diagonal terms, $\overline{\overline{\chi}}_{e}$ and $\overline{\overline{\chi}}_{m}$, define electric and magnetic susceptibilities. These susceptibilities  have two contributions. The intrinsic contribution, given by ${\overline{\overline \chi}}_{e_i} = \partial {\bf P}_i / \partial {\bf E} \; (= -\partial^2{\cal G}/\partial {\bf E}^2$), and ${\overline{\overline \chi}}_{m_i} = \partial {\bf M}_i / \partial {\bf B} \; (= -\partial^2{\cal G}/\partial {\bf B}^2$), and the toroidal contributions arising from toroidization. These last contributions are given by,
\begin{eqnarray}
{\overline{\overline \chi}}_{e_t} & = & \frac{\partial {\bf P}_t}{\partial {\bf E}} = \frac{\partial {\bf B} \times {\bf T}}{\partial {\bf E}} = {\bf B} \times \frac{\partial {\bf T}}{\partial {\bf E}}, \\
{\overline{\overline \chi}}_{m_t} & = &
\frac{\partial {\bf M}_t}{\partial {\bf B}} = \frac{\partial {\bf T} \times {\bf E}}{\partial {\bf B}} =  \frac{\partial {\bf T}}{\partial {\bf B}} \times {\bf E}.
\end{eqnarray}

A toroidal susceptibility can also be defined as,
\begin{eqnarray}
{\overline{\overline \chi}}_T = \partial {\bf T} / \partial {\bf G} = -\partial^2{\cal G}/\partial {\bf G}^2.
\end{eqnarray}
Neglecting the intrinsic contributions to the polarization and magnetization, this toroidal susceptibility can be expressed as,
\begin{eqnarray}
{\overline{\overline \chi}}_T
= \left[ {\bf E} \times \frac{1}{\frac{\partial {\bf T}}{\partial {\bf B}}\times {\bf E}}  - \frac{1}{\frac{\partial {\bf T}}{\partial {\bf E}} \times {\bf B} } \times {\bf B} \right]^{-1}
= \left[ {\bf E} \times \frac{1}{ {\overline{\overline\chi} }_{m_t} } +  \frac{1}{{\overline{\overline \chi}}_{e_t} }\times {\bf B}  \right]^{-1},
\end{eqnarray}
%
which shows that it is related to ${\overline{\overline \chi}}_{e_t}$ and ${\overline{\overline \chi}}_{m_t}$.

Cross terms define the magnetoelectric coefficients. Maxwell relations require that second derivatives of ${\cal G}$ are independent of the order in which they are performed. Therefore, this yields
\begin{eqnarray}
{\overline{\overline{\alpha}}}^T = {\overline{\overline{\alpha}}}.
\end{eqnarray}
Assuming that the whole magnetoelectric interplay arises from the existence of toroidization, the magnetoelectric coefficient is given by,
\begin{eqnarray}
{\overline{\overline{\alpha}}} = \frac{\partial {\bf P}_t}{\partial {\bf B}} = \frac{\partial {\bf T} \times {\bf B}} {\partial {\bf B}} =  \frac{\partial {\bf T}}{\partial {\bf B}} \times {\bf B} + {\bf T} \times \mathbb{I},
\end{eqnarray}
and
\begin{eqnarray}
{\overline{\overline{\alpha}}}^T = \frac{\partial {\bf M}_t}{\partial {\bf E}} =  \frac{\partial {\bf E} \times {\bf T}} {\partial {\bf E}} = {\bf E} \times \frac{\partial {\bf T}}{\partial {\bf E}}  + \mathbb{I} \times {\bf T},
\end{eqnarray}
where $\mathbb{I}$ is the identity tensor.
Notice that thermodynamic stability requires that $\xi$ is positive-definite. This implies that both ${\overline{\overline \chi}}_{e}$ and ${\overline{\overline \chi}}_{m}$ must be positive-definite and, ${\overline{\overline \chi}}_{e} \; {\overline{\overline \chi}}_{m} \geq \overline{\overline{\alpha}}^T \; \overline{\overline{\alpha}}$.

Thermal response is determined by the second order derivatives of ${\cal G}$ involving temperature. On the one hand, the heat capacity $C$ is given by
\begin{eqnarray}
\frac{\partial^2 {\cal G}}{\partial \tau^2} = \frac{C}{\tau}. \label{HeatCapacity}
\end{eqnarray}
Taking into account Maxwell relations, the derivatives involving temperature and fields satisfy,
\begin{eqnarray}
\frac{\partial^2 {\cal G}}{\partial \tau  \partial {\bf E}} = \frac{\partial^2 {\cal G}}{\partial {\bf E} \partial \tau} \Rightarrow \frac{\partial S}{\partial {\bf E}} = \frac{\partial {\bf P}}{\partial \tau},
\label{electrocaloric-response}
\end{eqnarray}
and
\begin{eqnarray}
\frac{\partial^2 {\cal G}}{\partial \tau  \partial {\bf B}} = \frac{\partial^2 {\cal G}}{\partial {\bf B} \partial \tau}  \Rightarrow \frac{\partial S}{\partial {\bf B}} = \frac{\partial {\bf M}}{\partial \tau} .
\label{magnetocaloric-response}
\end{eqnarray}
It can also be obtained that
\begin{eqnarray}
\frac{\partial^2 {\cal G}}{\partial \tau  \partial {\bf G}} = \frac{\partial^2 {\cal G}}{\partial {\bf G} \partial \tau}  \Rightarrow \frac{\partial S}{\partial {\bf G}} = \frac{\partial {\bf T}}{\partial \tau} .  \label{MaxwellToroidal}
\end{eqnarray}
These expressions determine the cross-response to electric or magnetic field and temperature. They are adequate for the study of thermal response  of the materials to applied external fields which are commonly denoted as caloric effects. These effects are quantified by the entropy change that occurs by isothermally applying or removing a given field, and the temperature change that results when the same field is applied or removed adiabatically. From a practical point of view, materials displaying large caloric effects are nowadays of great interest thanks to their potential use in energy harvesting, and particularly, in refrigeration applications \cite{Gutfleisch2011}. In general in ferroic and multiferroic materials, large caloric effects are expected in the vicinity of phase transitions to  ferroic and multiferroic phases due to the expected strong temperature dependence of thermodynamic properties \cite{Planes2014}. In the case of toroidal materials, caloric effects have been analyzed from a theoretical perspective in Ref. \cite{Castan2012}. The entropy change induced by application of an electric field, $(0 \rightarrow {\bf E})$, which quantifies the electrocaloric effect, can be obtained from integration of eq. (\ref{electrocaloric-response}) as,
\begin{eqnarray}
\Delta S(\tau, 0 \rightarrow {\bf E}) = \int_0^{{\bf E}} \frac{\partial {\bf P}}{\partial \tau} \cdot d{\bf E}.
\end{eqnarray}
Similarly the entropy change induced by application of a magnetic field, $(0 \rightarrow {\bf B})$, which quantifies the magnetocaloric effect, can be obtained from integration of eq. (\ref{magnetocaloric-response}) as,
\begin{eqnarray}
\Delta S(\tau, 0 \rightarrow {\bf B}) = \int_0^{{\bf B}} \frac{\partial {\bf M}}{\partial \tau} \cdot d{\bf B}.
\end{eqnarray}
Taking into account eqs. (\ref{P=Pi+Pt}) and (\ref{M=Mi+Mt}), the preceding entropy changes characterizing electrocaloric and magnetocaloric effects, can be, respectively, decomposed into two terms associated with intrinsic contributions and contributions arising from the toroidal moment. The intrinsic electro- and magnetocaloric terms are respectively,
\begin{eqnarray}
\Delta S_i(\tau, 0 \rightarrow {\bf E}) & = & \int_0^{{\bf E}} \frac{\partial {\bf P}_i}{\partial \tau} \cdot d{\bf E}, \\
\Delta S_i(\tau, 0 \rightarrow {\bf B}) & = & \int_0^{{\bf B}} \frac{\partial {\bf M}_i}{\partial \tau} \cdot d{\bf B}.
\end{eqnarray}
The contributions arising from the toroidal moment can be written in the form,
\begin{eqnarray}
\Delta S_t(\tau, 0 \rightarrow {\bf E}) & = & \int_0^{{\bf E}} \frac{\partial {\bf P}_t}{\partial \tau} \cdot d{\bf E} = \int_0^{{\bf E}} \left({\bf B} \times \frac{\partial {\bf T}}{\partial \tau} \right) \cdot d{\bf E},\\
\Delta S_t(\tau, 0 \rightarrow {\bf B}) & = & \int_0^{{\bf B}} \frac{\partial {\bf M}_t}{\partial \tau} \cdot d{\bf B} = \int_0^{{\bf B}} \left(\frac{\partial {\bf T}}{\partial \tau} \times {\bf E} \right) \cdot d{\bf B}.
\end{eqnarray}
A change of entropy can be isothermally induced by application of a toroidal field ${\bf G} = {\bf E} \times {\bf B}$. This entropy change characterizes the toroidocaloric effect, and when taken into account with eq. (\ref{MaxwellToroidal}), it can simply be expressed as,
\begin{eqnarray}
\Delta S(\tau, 0 \rightarrow {\bf G}) & = & \int_0^{{\bf G}} \frac{\partial {\bf T}}{\partial \tau} \cdot d{\bf G} = \int_0^{{\bf E}} \left({\bf B} \times \frac{\partial {\bf T}}{\partial \tau} \right) \cdot d{\bf E} + \int_0^{{\bf B}} \left(\frac{\partial {\bf T}}{\partial \tau} \times {\bf E} \right) \cdot d{\bf B} \label{S-toroidocal} \nonumber \\
& = & \Delta S_t(\tau, 0 \rightarrow {\bf E}) + \Delta S_t(\tau, 0 \rightarrow {\bf B}),
\end{eqnarray}
which shows that the toroidocaloric entropy change is simply the sum of the electrocaloric and magnetocaloric contributions associated with the toroidal moment as expected.

Similar expressions can be written for electrically and magnetically induced adiabatic temperature changes. The corresponding total change can be obtained by taking into account that from eqs. (\ref{HeatCapacity}), (\ref{electrocaloric-response}) and (\ref{magnetocaloric-response}), the constant entropy condition (adiabaticity in thermodynamic equilibrium) can be expressed as,
\begin{eqnarray}
\frac{C}{\tau} d \tau + \frac{\partial {\bf P}}{\partial \tau} \cdot d{\bf E}+ \frac{\partial {\bf M}}{\partial \tau} \cdot d{\bf B} = 0.
\end{eqnarray}
Therefore, the adiabatic temperature change induced by application of an electric field is given as,
\begin{eqnarray}
\Delta \tau (S, 0 \rightarrow {\bf E}) = \int_0^{{\bf E}} \frac{\tau}{C} \frac{\partial {\bf P}}{\partial \tau} \cdot d{\bf E},
\end{eqnarray}
and the adiabatic temperature change induced by application of a magnetic field as,
\begin{eqnarray}
\Delta \tau (S, 0 \rightarrow {\bf B}) = \int_0^{{\bf B}} \frac{\tau}{C} \frac{\partial {\bf M}}{\partial \tau} \cdot d{\bf B} .
\end{eqnarray}
On its turn, the adiabatic temperature change induced by application of a toroidal field is given by,
\begin{eqnarray}
\Delta \tau (S, 0 \rightarrow {\bf G}) = \int_0^{{\bf G}} \frac{\tau}{C} \frac{\partial {\bf T}}{\partial \tau} \cdot d{\bf T} = \int_0^{{\bf E}} \frac{\tau}{C} \left({\bf B} \times \frac{\partial {\bf T}}{\partial \tau} \right) \cdot d{\bf E} + \int_0^{{\bf B}} \frac{\tau}{C} \left(\frac{\partial {\bf T}}{\partial \tau} \times {\bf E} \right) \cdot d{\bf B}, \label{T-toroidocal}
\end{eqnarray}
where the last two terms in the right-hand side correspond to the sum of electrocaloric and magnetocaloric contributions associated with the toroidal moment. That is,
\begin{eqnarray}
\Delta \tau (S, 0 \rightarrow {\bf G}) = \Delta \tau_t (S, 0 \rightarrow {\bf E}) + \Delta \tau_t (S, 0 \rightarrow {\bf B}).
\end{eqnarray}
It is worth noting that both $\Delta S (\tau, 0 \rightarrow {\bf G})$ and $\Delta \tau (S, 0 \rightarrow {\bf G})$ vanish if either ${\bf E}$ or ${\bf B}$ is zero or if they are parallel. For the sake of simplicity let us consider that ${\bf E} = (E, 0, 0)$ and ${\bf B} = (0, B, 0)$. In this case, ${\bf G} = (0, 0, EB)$ and assuming electric and magnetic isotropy, ${\bf P} = (P, 0, 0)$, ${\bf M} = (0, M, 0)$ and ${\bf T} = (0, 0, T)$, and eqs. (\ref{S-toroidocal}) and (\ref{T-toroidocal}) are simply expressed as
\begin{eqnarray}
\Delta S(\tau, 0 \rightarrow EB)  =  \int_0^{EB} \frac{\partial T}{\partial \tau} d (EB) = B \int_0^{E} \frac{\partial T}{\partial \tau} d E + E \int_0^{B} \frac{\partial T}{\partial \tau} d B,
\end{eqnarray}
and
\begin{eqnarray}
\Delta \tau(S, 0 \rightarrow EB)  =  \int_0^{EB} \frac{\tau}{C} \frac{\partial T}{\partial \tau} d (EB) = B \int_0^{E} \frac{\tau}{C} \frac{\partial T}{\partial \tau} d E + E \int_0^{B} \frac{\tau}{C} \frac{\partial T}{\partial \tau} d B,
\end{eqnarray}
respectively.

\section{6. Landau and Ginzburg-Landau modelling and domains}

Phase transitons in ferroic and multiferroic materials are associated, as we have discussed above, with some symmetry change. This change is captured by an order parameter which is zero at temperatures above the transition and non-zero below it. Landau and Ginzburg-Landau theories provide reliable expressions of the free energy of the materials in the region of the transition in homogeneous and non-homogeneous cases, respectively. The approach is phenomenological in nature and its combination with the thermodynamics formalism provides a powerful method to study macroscopic and mesoscopic behaviour of ferroic and multiferroic materials. This approach enables one to relate measurable quantities to the input parameters of the theories that can be determined either from experiments or from first-principle calculations. In Landau theory the free energy is expressed as a series expansion of the order parameter. As this free energy must be invariant under the symmetry operations of the system, only those terms allowed by symmetry are included in the series expansion.

In ferrotoroidic materials toroidization is the primary order parameter but magnetization and polarization must also be included in the free energy. So far, few Landau models have been proposed to account for ferrotoroidal transition in specific materials. Sannikov \cite{Sannikov1998} already proposed a model to account for the anomalous behaviour of the component $\alpha_{32}$ of the magnetoelectric tensor near the cubic ($43m1'$) to the orthorrombic ($m' m2'$) phase transition in boracites.

Based on a group theoretic analysis Sannikov has provided a free energy for ferrotoroidic phase transitions in boracites \cite{Sannikov1998, Sannikov2007}.  It consists of a usual double well $F(T)=aT^2+bT^4$ in the toroidization $T$ and harmonic terms $cP^2$ and $dM^2$ in polarization and magnetization.  In addition, it
has symmetry allowed coupling terms between various components of the three dipolar vectors, namely of the form $P_iT_j^2$, $M_iT_j^3$ and the
trilinear coupling $P_iM_jT_k$.  The free energy is then analysed to obtain a phase diagram in terms of the free energy coefficients indicating the various ferrotoroidic phase transitions.  Note that Litvin has extended the symmetry analysis for ferroics to ferrotoroidic materials including the possible toroidic domains and domain walls \cite{Litvin08, LitvinActa, Litvin14}.  This analysis is very helpful in obtaining the free energy for ferrotoroidic phase transitions for any crystal symmetry.


Ederer and Spaldin, taking into account that the symmetries which allow for a macroscopic toroidal moment are the same that give rise to an antisymmetric component of the linear magnetoelectric effect tensor, proposed the simplest possible Landau free energy that describes a phase transition between a paratoroidic and a ferrotoroidic phase that includes the energies associated with the effect of electric and magnetic field on polarization and  magnetization, and their coupling to toroidization \cite{Ederer2007}. It has the following form,
\begin{eqnarray}
F(T, {\bf T}, {\bf P}, {\bf M}) & = & \frac{1}{2}A_0(\tau-\tau_c^0) T^2 +
\frac{1}{4} C T^4  + \frac{1}{2} \chi_p^{-1} P^2 +
\frac{1}{2}
\chi_m^{-1} M^2 \nonumber \\
& - & {\bf B} \cdot {\bf M} -{\bf E} \cdot {\bf P} +  \kappa {\bf T} \cdot ({\bf P} \times {\bf M}), \label{free-energy}
\end{eqnarray}
where $\chi_p$ and $\chi_m$ are electric and magnetic susceptibilities respectively, $A_0$ is the {\it toroidic stiffness}, and $C>0$ is the nonlinear toroidic coefficient. $\kappa$ measures the strength of the magnetoelectric coupling. The last term in the previous free energy represents the lowest possible order coupling term between the three order parameters consistent with the required space and time reversal symmetries.

Minimization of the free energy (\ref{free-energy}) with respect to polarization and magnetization provides the equilibrium values of polarization and magnetization. They are given by the following equations,
\begin{eqnarray}
{\bf P} = \chi_p {\bf E} - \chi_p \kappa ({\bf M} \times {\bf T}),
\end{eqnarray}
and
\begin{eqnarray}
{\bf M} = \chi_m {\bf B} - \chi_m \kappa ({\bf T} \times {\bf P}).
\end{eqnarray}
We can now solve these two equations assuming (for simplicity) that ${\bf E} =
(E,0,0)$ and ${\bf B} = (0,B,0)$ and, assuming isotropy,  ${\bf P} =(P,0,0)$, ${\bf M} = (0,M,0)$ and  ${\bf T} = (0,0,T)$.
We obtain:
\begin{eqnarray}
P = \chi_p E - \kappa \chi_p \chi_m B T + O(T^2) \simeq \chi_p E - \alpha B  \label{eqP}
\end{eqnarray}
and
\begin{eqnarray}
M = \chi_m H - \kappa \chi_p \chi _m E T + O(T^2) \simeq \chi_m B - \alpha E  , \label{eqM}
\end{eqnarray}
where we have neglected the nonlinear magnetoelectric effects in the above two equations.
The magnetoelectric coefficient $\alpha = \kappa\chi_p\chi_m T$ is a quadrilinear product of electric susceptibility, magnetic susceptibility, the coupling constant $\kappa$ and the toroidization.  Thus, either for $\kappa = 0$ or $\tau =0$ there is no magnetoelectric effect.

Substitution of $P$ (\ref{eqP}) and $M$ (\ref{eqM}) in the free energy
(\ref{free-energy}) gives the following general type of effective free energy:
\begin{eqnarray}
F_{e} & = & F_0(E, H) + \frac{1}{2} A_0 (\tau -\tau_c) T^2 + \frac{1}{3} \beta T^3 + \frac{1}{4} C T^4 + \lambda T ,
\label{eff-free-energy}
\end{eqnarray}
where,
\begin{eqnarray}
F_0 & = & -\frac{1}{2} \left(\chi_p E^2 + \chi_m B^2 \right) , \\
\tau_c & = & \tau_c^0 + \frac{\kappa^2}{A_0} \chi_p \chi_m [\chi_m B^2 + \chi_p E^2] ,
\label{Tc}\\
\beta & = & 3\kappa^3 \chi_m^2 \chi_p^2 EB = 3\kappa^2 \chi_m \chi_p \lambda, \label{B}\\
\lambda & = & \kappa  \chi_m  \chi_p EB .
\end{eqnarray}
The effective free energy (\ref{eff-free-energy}) corresponds to the free energy of a toroidal system subjected to an effective applied toroidal field $\lambda$ (proportional to $G$). Interestingly, the coefficient of the third order term, $\beta$, depends also on $\lambda$ (and thus, on $G$). When $G = 0$, $\beta=0$ then (\ref{eff-free-energy}) describes a paratoroidal-to-ferrotoroidal second-order phase transition. Under the application of a toroidal
field $G \neq 0$ (and $\beta \neq 0$), the transition becomes first-order for $|\lambda| > |\beta|^3 C^2/27$. It is worth pointing out that the addition of nonlinear terms in eqs. (\ref{eqP}) and (\ref{eqM}) would lead to higher order terms in the expansion (\ref{eff-free-energy}) that go beyond the minimal model. However, within the spirit of the Landau Theory, it is expected that such terms are not essential.

\begin{figure}[ht]
\centering
\includegraphics[scale=0.5]{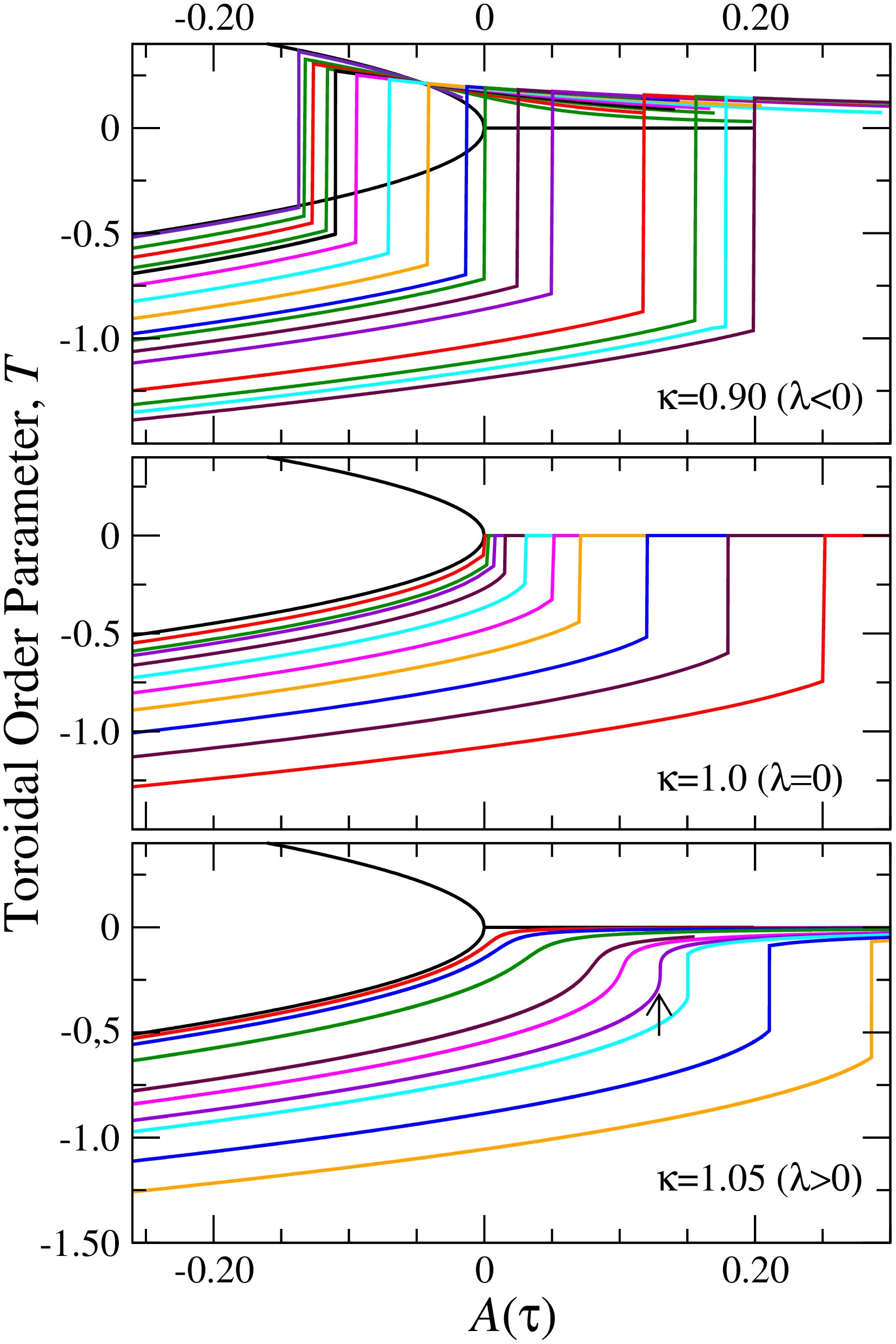}
\caption{Toroidal order parameter as a function of temperature (in arbitrary units) for three different values of the coupling parameter $\kappa=0.90$, $\kappa=1$ and $\kappa=1.05$ and selected values of the applied toroidal field $G$.  The arrow in the lower panel indicates the occurrence of the first order transition. Taken from \cite{Castan2012}.}
\label{Fig4}
\end{figure}

It is worth noting that the free energy (\ref{eff-free-energy}) does not include a term directly coupling ${\bf G}$ and ${\bf T}$, which is in agreement with the thermodynamic formulation developed in section 5\ref{Thermodynamics}. In Ref. \cite{Ederer2007}, however, ${\bf P}$, ${\bf M}$ and ${\bf T}$ were assumed to be independent order parameters and a term $-{\bf G} \cdot {\bf T}$ was included in the free energy. The same effective free energy (\ref{eff-free-energy}) is obtained, but in this case the parameter $\lambda$ is given by, $(\kappa \chi_m \chi_p -1)EB$. Similar results are obtained with this model which leads to a richer physics due to the fact that the $\lambda$ and $\beta$ terms can have different sign. The model has been used in \cite{Castan2012} in order to study toroidocaloric effects in ferrotoroidic materials. The entropy of the system can be obtained as,
\begin{eqnarray}
S(\tau, T, G) = - \frac{\partial F_e}{\partial \tau} =  - \frac{1}{2} A_0 T^2(\tau, G),
\end{eqnarray}
where $T (\tau, G)$ is the equilibrium value of the toroidal order parameter which is a solution of $ \partial F_e / \partial T = 0$. Then, the change of entropy isothermally induced by application of a toroidal field is obtained as,
\begin{eqnarray}
S(T, G=EH) - S(T, G = 0) =  - \frac{1}{2} A_0 [T^2(T, G=EH) - T^2(T, G =0)].
\end{eqnarray}
It is easy to show (see \cite{Planes2014}) that this expression coincides with the general thermodynamic expression (\ref{S-toroidocal}).  The variation of toroidization as a function of temperature for three different values of the coupling parameter ($\kappa$) is shown in Figure 4.
The corresponding isothermal entropy change ($\Delta S$) or the toroidocaloric effect is depicted in Figure 5.

\begin{figure}[ht]
\centering
\includegraphics[scale=0.5]{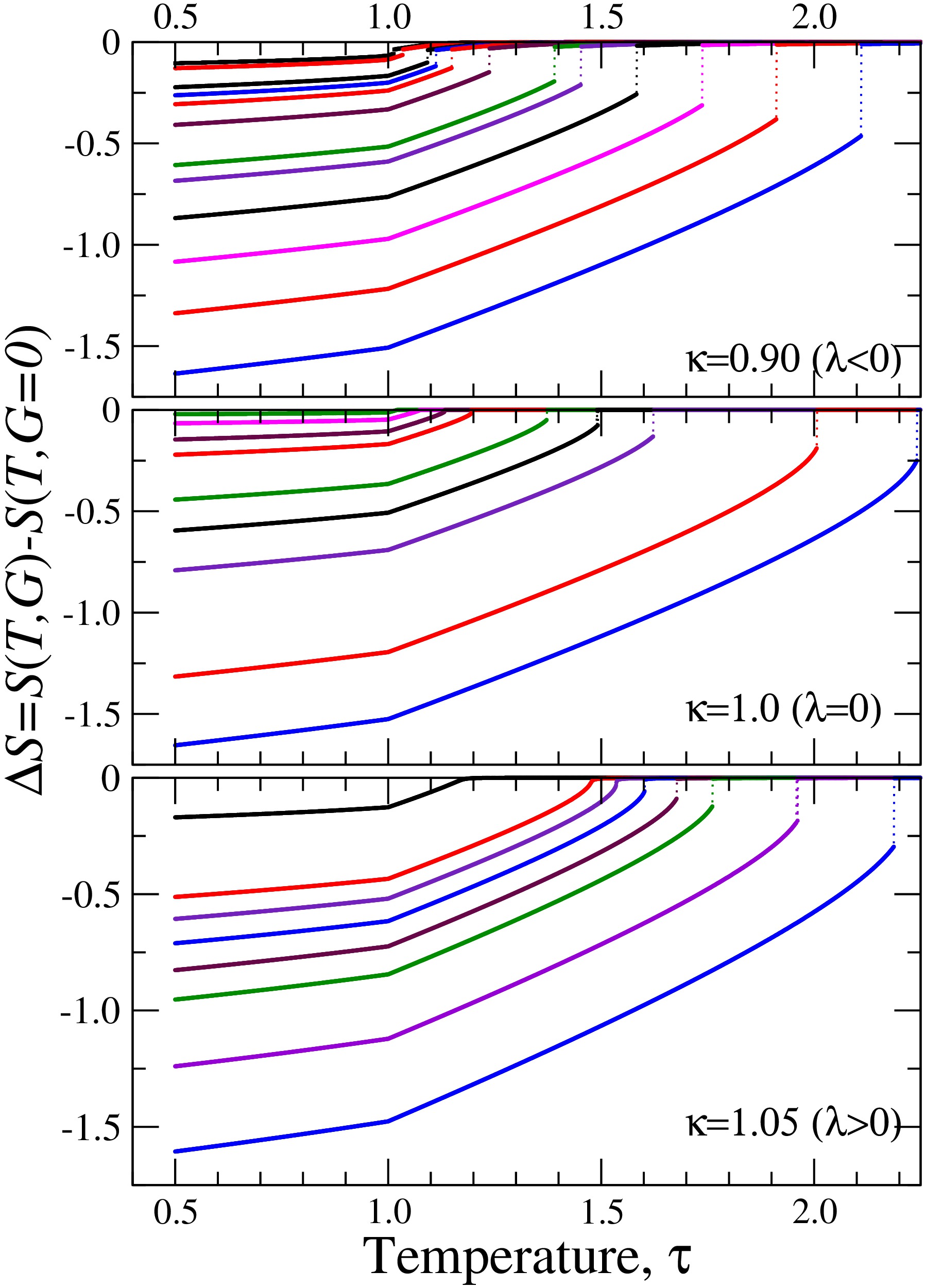}
\caption{Toroidocaloric effect, i.e. the isothermal change in entropy, as a function of transition temperature (in arbitrary units) and selected values of the applied toroidal field $G$ for three different values of the coupling parameter $\kappa=0.90$, $\kappa=1$ and $\kappa=1.05$. Taken from \cite{Castan2012}. }
\label{Fig5}
\end{figure}

The present Landau approach to ferrotoroidic materials can be generalized  by including symmetry allowed gradient term $(\nabla
{\bf T})^2$ \cite{Kopaev2009}, i.e. the Ginzburg term. This should allow one to study domains and domain walls in ferrotoroidic materials as observed in LiCo(PO$_4$)$_3$ using optical second harmonic generation techniques
\cite{VanAken2007}. With doping induced disorder in such materials we expect that novel phases such as toroidic tweed and toroidic glass should also
exist and remain to be observed experimentally with certainty \cite{Yamaguchi2013}. With symmetry allowed coupling of strain to toroidization, if we apply stress to such a crystal we
expect toroidoelastic effects, i.e. a change in toroidization with hydrostatic pressure or shear.  We expect that these important topics will be
explored in near future.


\section{7. Electric ferrotoroidics at the nanoscale}

We have already discussed in Section 3\ref{toroidal-moment} that no broken symmetry is associated with electric toroidization which is consistent with the fact the formation of electric moment vortex is forbidden in the thermodynamic limit. However, the situation can drastically change when the scale of the material decreases towards the nanoscale. It has been predicted that vortex structures can be stabilized below a certain critical temperature in both ferroelectric and ferromagnetic nanodots \cite{Dawber2003,Bader2006}. These zero-dimensional structures have been studied in detail from {\it ab initio} simulations of an effective hamiltonian under appropriate boundary conditions \cite{Naumov2004}. They numerically simulated Pb(Zr,Ti)O$_3$ ferroelectric nanodisks and nanorods under open circuit-like electric boundary conditions and found vortex states comprising electric dipoles that form a closed structure yielding spontaneous electric toroidal moment. Indeed, in recent years electrotoroidic behavior has been found experimentally \cite{Gruverman2008, McQuaid2011, Vasudevan2011, Nelson2011, Balke2012}.  Atomistic simulations of KTaO$_3$ seem to indicate an incipient ferrotoroidic response wherein quantum vibrations suppress the formation of polar vortices \cite{Prosan2009}. In addition, electrogyration or electric field induced optical rotation has been predicted in materials exhibiting electrotoroidic behavior \cite{Prosandeev2013}. They suggest that these nanostructures are potentially interesting for data storage applications.  Note that the effect of long-range elastic interactions on the electric toroidization in a ferroelectric nanoparticle has been considered in \cite{Wang2006}.

From a practical point of view, using these nanostructures based on electric toroidal moment as writing memory nanodevices is not straightforward since, in principle, these moments cannot be switched by standard methods as they are unaffected by applied electric fields.  However, several solutions have
been envisaged. In \cite{Prosandeev2006} it has been shown that vortices, both electric and magnetic, can be manipulated by inhomogeneous static fields.
 The coupling with elasticity enables controlling vortices by mechanical load, which gives rise to a rich temperature-stress phase diagram \cite{Chen2012}.
  Another possibility, considers nanodots with the shape of nanorings with an off-central hole. The interest in these objects relies on the fact that they are characterized by a transverse hypertoroidal moment, which is a polar vector and thus sensitive to an homogeneous applied electric field \cite{Pronsandeev2008}. Following similar ideas, it has been recently proposed \cite{Thorner2014} and numerically predicted that ferroelectric nanotori  can possess an homogeneous hypertoroidal moment as well as exhibit the coexistence of axial toroidal moment and hypertoroidal moment phases. In these nanoscale objects the hypertoroidal moment could be manipulated by an homogeneous applied electric field.  Note that as a different application, metamaterials based on the electric toroidal moment have been proposed \cite{Guo2012}.

\section{8. Conclusions}
With the recent emergence of magnetoelectric and multiferoic materials the fourth primary ferroic property, namely ferrotoroidicity, has gained special attention. In the present article we have developed a general thermodynamic framework for the study of phase transitions, domain walls and caloric effects within the context of Landau theory in ferrotoroidic materials such as LiCo(PO$_4$)$_3$.  Both magnetic and electric toroidal moments were considered although the latter can only exist in polar nanostructures. The generalization to hypertoroidal moments was also presented.  We discussed a variety of materials where toroidic order has been observed.  In the presence of sufficient disorder the other three primary ferroics exhibit glassy behaviour, namely as spin glass, relaxor ferroelectrics and strain glass \cite{Ren2009}.  Similarly, toroidal glass has been potentially observed as well in the study of dynamics of a linear magnetoelectric Ni$_{0.4}$Mn$_{0.6}$TiO$_3$ \cite{Yamaguchi2013}.  Presence of toroidal moments is also an indicator of magnetoelectric coupling in the material \cite{Ederer2009}.  Similarly, toroidal magnon excitations in multiferroics relate to magneto-optical effects \cite{Miyahara2014}. In this article we did not consider the fields and radiation from moving toroidal dipoles which is also an important area of research \cite{Ginzburg1985, Heras1998}.  Clearly, exploration of toroidal phenomena is a fertile area of research with a great potential for both fundamental science and device applications.  Indeed, toroidal metamaterials have been proposed, studied and experimentally realized \cite{Marinov2007, Kaelberer2010, Fan2013} including in double-ring \cite{Dong2012} and double-disk \cite{Li2014} structures.

\section{Acknowledgments}

This work received financial support from CICyT  (Spain), Project No. MAT2013-40590-P and was partially supported by the U.S. Department of Energy.

%
%

%
\end{document}